\documentstyle[preprint,prd,aps]{revtex}

\begin{document}
\draft
\preprint{\ \vbox{
\halign{&##\hfil\cr
       AS-ITP-2000-06\cr\cr}} } \vfil

\title{Effects of The Initial Hadron in $B\rightarrow J/\psi+X$}
\author{J.P. Ma}
\address{Institute of Theoretical Physics,\\
Academia Sinica, \\
P.O.Box 2735, Beijing 100080, China\\
e-mail: majp@itp.ac.cn}
\maketitle

\begin{abstract}
In the framework of HQET inclusive decays of b-flavored hadrons can be
handled as decays of the free b-quark approximately. We analyse the
correction to this approximation for the inclusive decays into a polarized $%
J/\psi$, the correction is characterized by two matrix elements defined in
HQET. For the $J/\psi$ we use NRQCD to parameterize the formation of the $c%
\bar c$ pair into $J/\psi$. Numerically the correction is remarkably large,
it can be at the level of $30\%$ and it is mainly due to the Fermi-motion of
the $b$-quark. With this correction we give a new determination for
combinations of two NRQCD matrix elements. \newline
\end{abstract}

\eject
\baselineskip=15pt

\vspace{-5mm}

\vskip20pt \narrowtext
Inclusive decay of a b-flavored hadron into a charmonium is an interesting
process in the aspect that two effective theories derived from QCD are used
to factorize nonperturbative effects. For the charmonium, because the
relative velocity $v_c$ between the $c$- and $\bar{c}$-quark inside the
charmonium is small in the rest-frame, one can use NRQCD\cite{BBL}to handle
the nonperturbative effect in the formation of the free $c\bar{c}$ pair into
the charmonium, while for the initial hadron one can employ an expansion in
the inverse of $m_b$--the b-quark mass, because $m_b$ is very large in
comparison with the nonperturbative energy-scale $\Lambda $. The expansion
can be done with the heavy quark effective theory(HQET)\cite{HQET}(for a
review, see e.g. \cite{Neu}). At the leading order the decay can be
considered as the decay of the free b-quark, hence the decay width is the
same for all initial hadrons. Corrections from higher orders can be
systematically added, they will depend on the type of initial hadrons. It is
the purpose of this work to study the corrections from the next-to-leading
order, i.e., the effect of initial hadrons.

It should be kept in mind that the factorization mentioned before is not
well established in comparison with the factorization for totally inclusive
decays and for inclusive semilepton-decay, in these decays one can use the
operator product expansion(OPE) to factorize nonperturbative effects. With
OPE the effect of the initial hadron in inclusive decays are studied, for
example, in \cite{BK,MW}. For the process considered here OPE can not be
used. However we can use the diagram expansion proposed in \cite{Poli} to
perform the factorization and assume that the factorization still holds,
i.e., the perturbatively calculated coefficients are free from infrared
divergences. We will use the expansion and work at the leading order of $%
\alpha _s$.

We consider the process: 
\begin{equation}
H_b(P)\rightarrow J/\psi (k,\lambda )+X
\end{equation}
where we denote the b-flavored hadron as $H_b$, which contains a b-quark.
The momenta are given in the brackets. $\lambda $ denotes the polarization
of the $J/\psi $, $\lambda =L$ is for the longitudinal polarization and $%
\lambda =T$ is for the transversal polarization. To analyze the decay three
expansions are used, they are expansion in $\alpha _s$, in $v_c$ and in the
inverse of $m_b$. In \cite{KLS,FH} the decay is studied at the leading order
of all expansions. At the leading order of $v_c$ and of the inverse of $m_b$
the one-loop correction is calculated in \cite{BMR} for the unpolarized $%
J/\psi $. The corrections from the next-to-leading order of $v_c$ is
analyzed in \cite{Ma} at the leading order of $\alpha _s$ for the polarized $%
J/\psi $, they can be very large. Because of lacking the detailed
information about the formation of the $c\bar{c}$ pair into a $J/\psi $,
i.e., the precise values of NRQCD matrix elements, a detailed prediction
seems impossible without help of some models. In this work we analyze the
corrections from the next-to-leading order of the inverse of $m_b$ to
complete the analysis of the correction from the next-to-leading order of
all expansions. By completing this analysis the last unknown effect in the
theory for the process can be estimated, in which nonperturbative effect is
represented by matrix elements defined in HQET and in NRQCD. In \cite{BSW}
models to account the effect are used to explain the $J/\psi $-spectrum.

The effective weak Hamiltonian for the decay is: 
\begin{eqnarray}
H_{{\rm eff}} &=&\frac{G_F}{\sqrt{2}}\sum_{q=s,d}\big\{V_{cb}V_{cq}^{*}[%
\frac 13C_{[1]}(\mu )\bar{c}\gamma ^\mu (1-\gamma _5)c\bar{q}\gamma _\mu
(1-\gamma _5)b  \nonumber \\
&\ &+C_{[8]}(\mu )\bar{c}T^a\gamma ^\mu (1-\gamma _5)c\bar{q}T^a\gamma _\mu
(1-\gamma _5)b]\big\}.
\end{eqnarray}
We neglected the contributions of QCD penguin operators in $H_{{\rm eff}}$. $%
T^a$($a=1,\cdots 8$) is SU(3) color-matrix. The coefficients $C_{[1]}$ and $C_{[8]}$
are related to the usual $C_{\pm }$ by 
\begin{equation}
C_{[1]}(\mu )=2C_{+}(\mu )-C_{-}(\mu ),\ \ \ C_{[8]}(\mu )=C_{+}(\mu )+C_{-}(\mu ).
\end{equation}
With the effective Hamiltonia a $c\bar{c}$ pair can not only be produced in
the color-singlet state, but also in the color-octet state. Although a $c%
\bar{c}$ pair in the color-octet state can be transmitted into a $J/\psi $
at higher orders in $v_c$ than a $c\bar{c}$ pair in the color-singlet state
does, its contributions to the decay rate are important, because the $c\bar{c%
}$ pair is more likely in the color-octet state than in the color-singlet
state by the fact $C_{[1]}(m_b):C_{[8]}(m_b)\approx 0.18:1$. We will take the
color-octet states into account and write the decay width as 
\begin{equation}
\Gamma _\lambda (J/\psi )=\Gamma _\lambda ^{(1)}+\Gamma _\lambda ^{(8)},
\end{equation}
where the index 1 or 8 stands for color singlet contributions or for
color-octet contributions, respectively. With the $H_{eff}$ and in the
approach of the diagram expansion at the leading order in $\alpha _s$, the
contributions to the decay width can be written: 
\begin{eqnarray}
\Gamma _\lambda ^{(1)} &=&\frac{G_F^2C^2_{[1]}}{18}|V_{cb}|^2\int d\Gamma \int
dx^4e^{iq\cdot x}\sum_X  \nonumber \\
&\cdot &\langle 0|\bar{c}(0)\gamma ^\mu (1-\gamma _5)c(0)|J/\psi +X\rangle
\langle J/\psi +X|c(x)\gamma ^\nu (1-\gamma _5)c(x)|0\rangle   \nonumber \\
&\cdot &\langle H_b|\bar{b}(0)\gamma _\mu (1-\gamma _5)\gamma \cdot q\gamma
_\nu (1-\gamma _5)b(x)|H_b\rangle ,
\end{eqnarray}
where we used $|V_{cd}|^2+|V_{cs}|^2\approx 1$. The integral $\int d\Gamma $
is 
\begin{equation}
\int d\Gamma =\int \frac{d^3k}{(2\pi )^3}\frac{d^4q}{(2\pi )^4}2\pi \delta
(q^2).
\end{equation}
In the above equations we neglect the mass of light quarks and take the
nonrelativistic normalization for $H_b$- and $J/\psi $-state. Similarly the
color-octet contribution is 
\begin{eqnarray}
\Gamma _\lambda ^{(8)} &=&\frac{G_F^2C^2_{[8]}}2|V_{cb}|^2\int d\Gamma \int
dx^4e^{iq\cdot x}\sum_X  \nonumber \\
&\cdot &\langle 0|\bar{c}(0)T^a\gamma ^\mu (1-\gamma _5)c(0)|J/\psi
+X\rangle \langle J/\psi +X|c(x)T^b\gamma ^\nu (1-\gamma _5)c(x)|0\rangle  
\nonumber \\
&\cdot &\langle H_b|\bar{b}(0)T^a\gamma _\mu (1-\gamma _5)\gamma \cdot
qT^b\gamma _\nu (1-\gamma _5)b(x)|H_b\rangle .
\end{eqnarray}

In both contributions the dependence on the initial hadron appears in a
matrix element, which is given by: 
\begin{equation}
T_{ij}(x)=\langle H_b|\bar{b}_j(0)b_i(x)|H_b\rangle ,
\end{equation}
where the indices $i,j$ stand for color- and Dirac indices and the average of
the spin is implied if $H_b$ has a non-zero spin. This matrix element
represents the nonperturbative effects related to the initial hadron, it can
be expanded in the inverse of $m_b$ in the framework of HQET. The expansion
is performed to expand the Dirac fields $\bar{b}(0)$ and $b(x)$ with the
fields in HQET. We use the expansion proposed in \cite{BKP}, which is
slightly different than the standard approach in \cite{HQET}. However, the
final results are same by using these two expansions, if one correctly takes
the normalization of the state in HQET into account. The Dirac field can be
expanded as: 
\begin{eqnarray}
b(x) &=&e^{-im_bv\cdot x}\left\{ 1+\frac 1{2m_b}i\gamma \cdot D_T+\frac 1{%
4m_b^2}v\cdot D\gamma \cdot D_T-\frac 1{8m_b^2}(\gamma \cdot D_T)^2\right\}
h(x)+{\cal O}(\frac 1{m_b^3})  \nonumber \\
&&+({\rm terms\ for\ anti-quark}),
\end{eqnarray}
where $v^\mu $ is the velocity of $H_b$, $D^\mu $ is the covariant
derivative $\partial ^\mu +igG^\mu (x)$. $D_T^\mu $ is defined as 
\begin{equation}
D_T^\mu =D^\mu -v^\mu v\cdot D.
\end{equation}
$h(x)$ is the field of HQET and its equation of motion reads: 
\begin{equation}
\left\{ iv\cdot D-\frac 1{2m_b}(\gamma \cdot D_T)^2\right\} h(x)=0+{\cal O}(%
\frac 1{m_b^2}).
\end{equation}
With the expansion in Eq.(9) the operator $N$ for the number of b-quarks is 
\begin{equation}
N=\int d^3x\bar{h}(x)h(x).
\end{equation}
Using the expansion in Eq.(9) the matrix element $T_{ij}(x)$ can be written
as 
\begin{eqnarray}
T_{ij}(x) &=&e^{-im_bv\cdot x}\big\{ \langle H_b|\bar{h}_j(0)h_i(x)|H_b%
\rangle +\frac 1{2m_b}\langle H_b|\bar{h}_j(0)(i\gamma \cdot D_Th(x))_i-i%
\overline{\gamma \cdot D_Th}_j(0)h_i(x)|H_b\rangle   \nonumber \\
&&+\frac 1{4m_b^2}\langle H_b|\bar{h}_j(0)(v\cdot D\gamma \cdot D_Th(x))_i+%
\overline{v\cdot D\gamma \cdot D_Th}_j(0)h_i(x)+\overline{\gamma \cdot D_Th}%
_j(0)(\gamma \cdot D_Th(x))_i  \nonumber \\
&&\ \ -\frac 12\bar{h}_j(0)((\gamma \cdot D_T)^2h(x))_i-\frac 12\overline{%
(\gamma \cdot D_T)^2h}_j(0)h_i(x)|H_b\rangle \big\} +{\cal O}(\frac 1{m_b^3}%
).
\end{eqnarray}
The dominant $x$-dependence of $T_{ij}(x)$ is $\exp (-im_bv\cdot x)$, while
the $x$-dependence of matrix elements with $h(x)$ is controlled by the
energy scale $\Lambda $, which is much smaller than $m_b$, we can expand
this $x$-dependence in $x$. In this expansion we obtain a tower of local
operators $O_n$. The equation of motion
Eq.(11) should be used to eliminate redundant operators. The contribution
from a given operator $O_n$ to the decay width will be suppressed by the
power $\Lambda^{(d_n-3)}$, where $d_n$ is the dimeneion of 
the operator $O_n$. For example, the first matrix element in Eq.(13)
is expanded as 
\begin{eqnarray}
\langle H_b|\bar{h}_j(0)h_i(x)|H_b\rangle  &=&\langle H_b|\bar{h}%
_j(0)h_i(0)|H_b\rangle +x_\mu \langle H_b|\bar{h}_j(0)\partial ^\mu
h_i(0)|H_b\rangle   \nonumber \\
&&+\frac 12x_\mu x_\nu \langle H_b|\bar{h}_j(0)\partial ^\mu \partial ^\nu
h_i(0)|H_b\rangle +\cdots .
\end{eqnarray}
With the Lorentz covariance and properties of the field $h(x)$, the matrix
element can be written 
\begin{eqnarray}
\langle H_b|\bar{h}_j(0)h_i(x)|H_b\rangle  &=&\frac 1{12}(1+\gamma \cdot
v)_{ij}\langle H_b|\bar{h}h|H_b\rangle +\frac 1{12}v\cdot x(1+\gamma \cdot
v)_{ij}\langle H_b|\bar{h}v\cdot \partial h|H_b\rangle   \nonumber \\
&&+\frac 1{72}(1+\gamma \cdot v)_{ij}x_\mu x_\nu \big[ g^{\mu \nu }\langle
H_b|\bar{h}\partial _T^2h|H_b\rangle   \nonumber \\
&&-v^\mu v^\nu \langle H_b|\bar{h}(\partial _T^2-3(v\cdot \partial
)^2h|H_b\rangle \big] +\cdots .
\end{eqnarray}
Because of Eq.(12) the first matrix element in the r.h.s. of the above
equation is 1: 
\begin{equation}
\langle H_b|\bar{h}h|H_b\rangle =1.
\end{equation}
This results in that the decay of $H_b$ at the leading order of $m_b^{-1}$
is the decay of the free b-quark. Similarly we can expand the other matrix
elements in Eq.(13) and we obtain: 
\begin{eqnarray}
T_{ij}(x) &=&e^{-im_bv\cdot x}\big\{ \frac 1{12}(1+\gamma \cdot v)_{ij}+%
\frac{ix^\mu }{2m_b}\big[ \frac 1{12}(1+\gamma \cdot v)_{ij}v_\mu   \nonumber
\\
&&+\frac 1{18}(\gamma _\mu -v_\mu \gamma \cdot v)\big] \langle H_b|\bar{h}%
(\gamma \cdot D_T)^2h|H_b\rangle +\frac 1{72}x^\mu x^\nu (g_{\mu \nu }-v_\mu
v_\nu )  \nonumber \\
&&\cdot (1+\gamma \cdot v)_{ij}\langle H_b|\bar{h}D_T^2h|H_b\rangle -\frac 1{%
24}\langle H_b|\bar{h}(\gamma \cdot D_T)^2h|H_b\rangle \big\} +\cdots 
\end{eqnarray}
where we have replaced the derivative $\partial _\mu $ with the covariant
one $D_\mu $ and the equation of motion was used to eliminate the redundant
operators. The $\cdots $ stand for contributions from operators with the
dimension higher than 5, which will lead to contributions at the order
higher than $\Lambda ^2$.

For the chamonium $J/\psi$ we will work at the leading order in $v_c$ and we
have $M_{J/\psi}=2m_c$. To expand the Dirac field $c(x)$ with the field of
NRQCD, we make a Lorentz boost to transform the relevant matrix element into
the $J/\psi$-rest frame and then make the expansion. We obtain for the
color-singlet part: 
\begin{eqnarray}
&& \sum_X \langle 0\vert \bar c(0)\gamma^\mu (1-\gamma_5) c(0)\vert J/\psi
+X \rangle \langle J/\psi+X\vert c(x)\gamma^\nu (1-\gamma_5) c(x) \vert
0\rangle  \nonumber \\
&& = e^{ik\cdot x} \frac{M_{J/\psi}}{k^0} \varepsilon^\mu(\lambda)
(\varepsilon^\nu(\lambda))^* \frac{1}{3} \langle O^{J/\psi}_1 (^3S_1)
\rangle +{\cal O} (v_c^2),
\end{eqnarray}
where the factor $M_{J/\psi}/k^0$ is due to the boost. The definition of the
matrix element $\langle O^{J/\psi}_1 (^3S_1) \rangle$ can be found in \cite
{BBL}.

With Eq.(17) and Eq.(18) we can perform the $x$-intergration in Eq.(5).
After the integration we obtain a $\delta$-function for the momentum
conservation, and also derivatives of the $\delta $-function like 
\begin{equation}
\frac \partial {\partial q_\mu }(2\pi )^4\delta ^4(m_bv-q-k).
\end{equation}
Such derivatives can be eliminated by partial integrations when the
integration of the phase-space is performed. For example, the contribution
related the above derivative can be written: 
\begin{equation}
\int d^4q\delta (q^2)f(q)\frac \partial {\partial q_\mu }\delta
^4(m_bv-q-k)=-(2q^\mu \delta ^{\prime }(q^2)f(q)+\delta (q^2)\frac \partial {%
\partial q_\mu }f(q))|_{q=m_bv-k},
\end{equation}
where $\delta ^{\prime }(x)=\frac d{dx}\delta (x)$. Performing these
integrations the final results for the color-singlet contribution can be
obtained. To present our results we write: 
\begin{eqnarray}
\langle H_b|\bar{h}D_T^2h|H_b\rangle  &=&-\mu _\pi ^2,  \nonumber \\
\langle H_b|\bar{h}(\gamma \cdot D_T)^2h|H_b\rangle  &=&\langle H_b|\bar{h}%
D_T^2h|H_b\rangle +\langle H_b|\bar{h}\frac 12gG_{\mu \nu }\sigma ^{\mu \nu
}h|H_b\rangle +\cdots   \nonumber \\
&=&-\mu _\pi ^2+\mu _g^2+\cdots ,
\end{eqnarray}
where $\cdots $ stand for higher-order contributions. The color-singlet
contribution $\Gamma _\lambda ^{(1)}$ can be written as 
\begin{equation}
\Gamma _\lambda ^{(1)}=\frac{m_b^3G_F^2C_{[1]}^2|V_{cb}|^2}{432\pi m_c}\langle
O_1^{J/\psi }(^3S_1)\rangle \cdot \left\{ F_\lambda (y)+\frac{\mu _\pi ^2}{%
m_b^2}H_\lambda (y)+\frac{\mu _g^2}{m_b^2}G_\lambda (y)\right\} 
\end{equation}
where $y=4m_c^2/m_b^2$ and the functions are: 
\begin{eqnarray}
F_L(y) &=&(1-y)^2,  \nonumber \\
F_T(y) &=&y(1-y)^2,  \nonumber \\
G_L(y) &=&\frac 16(7-4y+7y^2-10y^3),  \nonumber \\
G_T(y) &=&\frac 16y(11-26y+15y^2),  \nonumber \\
H_L(y) &=&\frac 16(-53+4y-y^2+26y^3),  \nonumber \\
H_T(y) &=&\frac 16y(-11+14y-27y^2).
\end{eqnarray}
From the above results the effect of the initial hadron is represented by
two matrix elements defined in HQET. It should be noted that the effect is
not suppressed by the power of $m_b^{-1}$, but rather by the power of $%
(m_b^2-4m_c^2)^{-1}$. This can be seen by comparing the coefficients given
above.

For the color-octet contribution we expand the the color-octet matrix
element related to $J/\psi $ with NRQCD fields. The result at the leading
order of $v_c$ is: 
\begin{eqnarray}
&&\sum_X\langle 0|\bar{c}(0)T^a\gamma ^\mu (1-\gamma _5)c(0)|J/\psi
+X\rangle \langle J/\psi +X|c(x)T^b\gamma ^\nu (1-\gamma _5)c(x)|0\rangle  
\nonumber \\
&=&e^{ik\cdot x}\frac{M_{J/\psi }}{k^0}\frac 1{24}\delta _{ab}\big\{ \frac{%
k^\mu k^\nu }{M_{J/\psi }}\langle O_8^{J/\psi }(^1S_0)\rangle +\varepsilon
^\mu (\lambda )(\varepsilon ^\nu (\lambda ))^{*}\langle O_8^{J/\psi
}(^3S_1)\rangle   \nonumber \\
&&+(-g^{\mu \nu }+\frac{k^\mu k^\nu }{M_{J/\psi }}-\varepsilon ^\mu (\lambda
)(\varepsilon ^\nu (\lambda ))^{*})\frac 1{m_c^2}\langle O_8^{J/\psi
}(^3P_1)\rangle \big\} +{\cal O}(v_c^6).
\end{eqnarray}
At this order, the color-octet $c\bar{c}$ pair with the quantum number $^1S_0
$, $^3S_1$ and $^3P_1$ can form a $J/\psi $ through emission or absorption
of soft gluons, the probability is at order of $v_c^4$. Perform similar
calculations as for the color-singlet contribution we obtain the color-octet
contribution: 
\begin{equation}
\Gamma _\lambda ^{(8)}=\frac{m_b^3G_F^2C_{[8]}^2|V_{cb}|^2}{288\pi m_c}\left\{
F_\lambda ^{(8)}(y)+\frac{\mu _\pi ^2}{m_b^2}H_\lambda ^{(8)}(y)+\frac{\mu
_g^2}{m_b^2}G_\lambda ^{(8)}(y)\right\} ,
\end{equation}
where the functions are: 
\begin{eqnarray}
F_L^{(8)}(y) &=&(1-y)^2(\langle O_8^{J/\psi }(^1S_0)\rangle +\langle
O_8^{J/\psi }(^3S_1)\rangle )+2y(1-y)^2\frac 1{m_c^2}\langle O_8^{J/\psi
}(^3P_1)\rangle ,  \nonumber \\
F_T^{(8)}(y) &=&(1-y)^2(\langle O_8^{J/\psi }(^1S_0)\rangle +y\langle
O_8^{J/\psi }(^3S_1)\rangle )+(1-y)^2(1+y)\frac 1{m_c^2}\langle O_8^{J/\psi
}(^3P_1)\rangle ,  \nonumber \\
G_L^{(8)}(y) &=&\frac 16((7-12y+15y^2-10y^3)\langle O_8^{J/\psi
}(^1S_0)\rangle +(7-4y+7y^2-10y^3)\langle O_8^{J/\psi }(^3S_1)\rangle  
\nonumber \\
&&+2(-1-14y+15y^2)\frac 1{m_c^2}\langle O_8^{J/\psi }(^3P_1)\rangle ), 
\nonumber \\
G_T^{(8)}(y) &=&\frac 16((7-12y+15y^2-10y^3)\langle O_8^{J/\psi
}(^1S_0)\rangle +y(11-26y+15y^2)\langle O_8^{J/\psi }(^3S_1)\rangle  
\nonumber \\
&&+(7-17y+5y^2+5y^3)\frac 1{m_c^2}\langle O_8^{J/\psi }(^3P_1)\rangle ), 
\nonumber \\
H_L^{(8)}(y) &=&\frac 16((-53+12y-9y^2+26y^3)\langle O_8^{J/\psi
}(^1S_0)\rangle +(53+4y-y^2+26y^3)\langle O_8^{J/\psi }(^3S_1)\rangle  
\nonumber \\
&&+2y(1+2y-27y^2)\frac 1{m_c^2}\langle O_8^{J/\psi }(^3P_1)\rangle ), 
\nonumber \\
H_T^{(8)}(y) &=&\frac 16((-53+12y-9y^2+26y^3)\langle O_8^{J/\psi
}(^1S_0)\rangle +y(-11+14y-27y^2)\langle O_8^{J/\psi }(^3S_1)\rangle  
\nonumber \\
&&+(-53+17y-11y^2-y^3)\frac 1{m_c^2}\langle O_8^{J/\psi }(^3P_1)\rangle ).
\end{eqnarray}

With the results given above we complete our analysis for the effect of the
initial hadron. As mentioned before, because the NRQCD matrix elements,
especially the color-octet matrix elements, are not known precisely and also
there are possibly large relativistic corrections for charmonium, we can not
give a precise prediction for the decay width. However, the effect of the
initial hadron can be estimated numerically. For this we take $m_b=4.8$GeV
and $m_c=1.5$GeV. We obtain for the color singlet contribution: 
\begin{eqnarray}
\Gamma _T^{(1)} &=&\frac{m_b^3G_F^2C_{[1]}^2|V_{cb}|^2}{432\pi m_c}\langle
O_1^{J/\psi }(^3S_1)\rangle \cdot \left\{ 0.15-0.63\frac{\mu _\pi ^2}{m_b^2}%
+0.20\frac{\mu _g^2}{m_b^2}\right\} ,  \nonumber \\
\Gamma _L^{(1)} &=&\frac{m_b^3G_F^2C_{[1]}^2|V_{cb}|^2}{432\pi m_c}\langle
O_1^{J/\psi }(^3S_1)\rangle \cdot \left\{ 0.37-8.34\frac{\mu _\pi ^2}{m_b^2}%
+0.98\frac{\mu _g^2}{m_b^2}\right\} .
\end{eqnarray}
It is remarkable that the coefficient in the front of $\mu _\pi ^2$ is very
large for $\Gamma _L^{(1)}$ in comparison with others. This indicates that
the effect of the initial hadron may be substantial. Taking $H_b=B$ as an
example, the parameter $\mu _g^2(B)$ can be determined by the mass splitting
between $B$ and $B^{*}$\cite{Vol}, it gives $\mu _g^2(B)\approx 0.36{\rm GeV}%
^2$, while $\mu _\pi ^2(B)$ can be determined from the QCD sum rules\cite{BB}
and from an analysis of spectroscopy of heavy hadrons\cite{Neu1}. These
determinations give a numerical range from $0.3{\rm GeV}^2$ to $0.54{\rm GeV}%
^2$ for $\mu _\pi ^2(B)$. There is also a constraint for $\mu _\pi ^2(B)$, $%
\mu _\pi ^2(B)\le\mu _g^2(B)$\cite{Vol1}. We take the value $\mu _\pi
^2(B)\approx \mu _g^2(B)$ to estimate the effect of the initial $B$. With these
values, the effect in $\Gamma _T^{(1)}$ is at $4\%$ level, which is
negligible, while the effect in $\Gamma _L^{(1)}$ is to reduce the width
at $30\%$ level in comparison with $\Gamma _L^{(1)}$ at the leading order, 
which is significant. Taking the
same quark masses we obtain for the color-octet contribution: 
\begin{eqnarray}
\Gamma _T^{(8)} &=&\frac{m_b^3G_F^2C_{[8]}^2|V_{cb}|^2}{288\pi m_c}\cdot \big\{ %
(0.37-8.02\frac{\mu _\pi ^2}{m_b^2}+0.67\frac{\mu _g^2}{m_b^2})\langle
O_8^{J/\psi }(^1S_0)\rangle   \nonumber \\
&&\ \ \ \ +(0.15-0.63\frac{\mu _\pi ^2}{m_b^2}+0.20\frac{\mu _g^2}{m_b^2}%
)\langle O_8^{J/\psi }(^3S_1)\rangle   \nonumber \\
&&\ \ \ \ +(0.52-8.02\frac{\mu _\pi ^2}{m_b^2}+0.23\frac{\mu _g^2}{m_b^2})%
\frac 1{m_c^2}\langle O_8^{J/\psi }(^3P_1)\rangle \big\}  \nonumber \\
\Gamma _L^{(8)} &=&\frac{m_b^3G_F^2C_{[8]}^2|V_{cb}|^2}{288\pi m_c}\cdot \big\{ %
(0.37-8.02\frac{\mu _\pi ^2}{m_b^2}+0.67\frac{\mu _g^2}{m_b^2})\langle
O_8^{J/\psi }(^1S_0)\rangle   \nonumber \\
&&\ \ \ \ +(0.37-8.34\frac{\mu _\pi ^2}{m_b^2}+0.98\frac{\mu _g^2}{m_b^2}%
)\langle O_8^{J/\psi }(^3S_1)\rangle   \nonumber \\
&&\ \ \ \ +(0.29-0.30\frac{\mu _\pi ^2}{m_b^2}+0.54\frac{\mu _g^2}{m_b^2})%
\frac 1{m_c^2}\langle O_8^{J/\psi }(^3P_1)\rangle \big\}.
\end{eqnarray}
In the color-octet contribution there are also large numbers in the front of 
$\mu _\pi ^2$ in different production channels. This results in that the
corrections are significant. We also take $B$-meson as an example and we
find: 
\begin{eqnarray}
\Gamma _T^{(8)}(B) &\approx &\frac{m_b^3G_F^2C_{[8]}^2|V_{cb}|^2}{288\pi m_c}%
\cdot \big\{ (0.37-0.11)\langle O_8^{J/\psi }(^1S_0)\rangle   \nonumber \\
&&+(0.15-0.007)\langle O_8^{J/\psi }(^3S_1)\rangle +(0.52-0.12)\frac 1{m_c^2}%
\langle O_8^{J/\psi }(^3P_1)\rangle \big\}  \nonumber \\
\Gamma _L^{(8)}(B) &\approx &\frac{m_b^3G_F^2C_{[8]}^2|V_{cb}|^2}{288\pi m_c}%
\cdot \big\{ (0.37-0.11)\langle O_8^{J/\psi }(^1S_0)\rangle   \nonumber \\
&&+(0.37-0.11)\langle O_8^{J/\psi }(^3S_1)\rangle +(0.29-0.004)\frac 1{m_c^2}%
\langle O_8^{J/\psi }(^3P_1)\rangle \big\},
\end{eqnarray}
where the second number in $(\cdots )$ is for the corrections. From these
numbers one can see that the corrections are large, especially in the $^1S_0$
channel, which can be at $30\%$ level. The large corrections arise in both
color-singlet and color-octet contributions from terms with ${\mu _\pi ^2}
$, that indicates that the Fermi-motion of the $b$-quark inside $B$ affects
the decay width substantially.

If we collect all known results to compare the experimental result\cite{exp} 
\begin{eqnarray}
Br(B\rightarrow J/\psi +X)_{{\rm exp}} &=&0.80\pm 0.08\%,  \nonumber \\
Br(B\rightarrow \psi ^{\prime }+X)_{{\rm exp}} &=&0.34\pm 0.05\%,
\end{eqnarray}
we have serious problems with the one-loop QCD correction and with the
relativistic correction. With the one-loop correction the color-singlet
contribution becomes negative\cite{BMR}. The relativistic correction is
analyzed in \cite{Ma} and it can have large effect, but the detailed size is
unknown because there are four unknown matrix elements of NRQCD. If we
neglect it we have: 
\begin{eqnarray}
Br(B\rightarrow \psi +X) &=&0.073\cdot 10^{-2}\langle O_1^\psi
(^3S_1)\rangle   \nonumber \\
&&+0.19\langle O_8^\psi (^3S_1)\rangle +0.33\langle O_8^\psi (^1S_0)\rangle +%
\frac{0.34}{m_c^2}\langle O_1^\psi (^3P_1)\rangle   \nonumber \\
&&-\big\{ 0.068\cdot 10^{-2}\langle O_1^\psi (^3S_1)\rangle +0.071\langle
O_8^\psi (^1S_0)\rangle   \nonumber \\
&&+0.027\langle O_8^\psi (^3S_1)\rangle +\frac{0.053}{m_c^2}\langle O_1^\psi
(^3P_1)\rangle \big\},
\end{eqnarray}
where we use $\psi $ to denote $J/\psi $ or $\psi ^{\prime }$. In the first
two lines there are the results from the leading order in the $m_b^{-1}$%
-expansion. The numbers are from \cite{BMR} and they contain one-loop QCD
corrections. They are slightly modified due to the $m_b^{-2}$ corrections in
the semiletonic decay width. It should be noted that the color-singlet
contribution is improved by adding certain contributions from higher orders
and with the improvement it becomes positive. The results of this work are
given in the last two lines. With the corrections the contributions from all
channels become smaller than those without the corrections. If we take 
a larger value of $\mu_{\pi}^2(B)$, the color-singlet contribution becomes 
negative. Among the four
matrix elements in Eq.(31) the color-single one is calculated with potential
models\cite{EQ} and with lattice QCD\cite{BSK}, whose value is\cite{EQ} 
\begin{eqnarray}
\langle O_1^{J/\psi }(^3S_1)\rangle  &=&1.16{\rm GeV}^3,  \nonumber \\
\langle O_1^{\psi ^{\prime }}(^3S_1)\rangle  &=&0.76{\rm GeV}^3.
\end{eqnarray}
The matrix element $\langle O_8^\psi (^1S_0)\rangle $ is determined by
experiments at $Tevatron$ and at $Hera$\cite{BF,CL,BK1,KK}, but the
uncertainty can be large. We take the value: 
\begin{eqnarray}
\langle O_8^{J/\psi }(^3S_1)\rangle  &=&1.06\cdot 10^{-2}{\rm GeV}^3, 
\nonumber \\
\langle O_8^{\psi ^{\prime }}(^3S_1)\rangle  &=&0.44\cdot 10^{-2}{\rm GeV}^3.
\end{eqnarray}
The other two are not well determined, only certain combination of them is
known with large uncertainty. With the experimental results in Eq.(30) we
can determine a combination of them: 
\begin{eqnarray}
\langle O_8^{J/\psi }(^1S_0)\rangle +\frac{1.13}{m_c^2}\langle O_1^{J/\psi
}(^3P_1)\rangle  &=&2.4\cdot 10^{-2}{\rm GeV}^3,  \nonumber \\
\langle O_8^{\psi ^{\prime }}(^1S_0)\rangle +\frac{1.13}{m_c^2}\langle
O_1^{\psi ^{\prime }}(^3P_1)\rangle  &=&1.0\cdot 10^{-2}{\rm GeV}^3.
\end{eqnarray}
Comparing the combinations determined without the corrections\cite{BMR} the
change is significant, the combinations becomes $80\%$ larger than those
without the corrections and they are closer to those determined from other
experiments\cite{BF,CL,BK1,KK}. However, the determination of the
combinations should be regarded as a rough one because possibly large
corrections from higher orders in $v_c$ are neglected.

To summarize: We have analyzed the effect of the initial hadron in the
process $H_b\rightarrow \psi +X$. With our results the three expansions used
for the process are all completed at the next-to-leading order. The effect
of the initial hadron can be at $30\%$ level for $B\rightarrow \psi +X$ and
it reduces the decay width. The effect is mainly due to the Fermi-motion of
the $b$-quark inside the initial hadron. Including this effect we have
determined combinations of two NRQCD matrix elements, which are $80\%$
larger than those without the effect and they are closer to those determined
in other experiment. However, the results for the combinations should be
taken with caution because neglected effects can be large.

\vskip20pt \noindent
{\bf Acknowledgment:} This work is supported by National Science Foundation
of P.R. China and by the Hundred Yonng Scientist Program of Sinica Academia
of P.R.China.

\vskip15pt

\vfil\eject

\end{document}